\documentclass[sigconf]{acmart}

\microtypesetup{expansion=false}

\usepackage{multirow}
\usepackage{subcaption}
\usepackage[most]{tcolorbox}
\usepackage{enumitem}
\AtBeginDocument{%
  \captionsetup[figure]{font=small}
}

\setcopyright{none}
\acmConference{}{}{}
\acmBooktitle{}
\acmISBN{}
\acmDOI{}
\renewcommand\footnotetextcopyrightpermission[1]{}

\begin{document}

\title[TrackFlood: Relocating Latency Attacks to Real-Time Trackers]{TrackFlood: Relocating Latency Attacks from NMS-Free Detectors to Real-Time Trackers}
\author{Zonghua Gu, Julian Singh-Smith, Junlin Liao}
\affiliation{%
  \institution{Hofstra University}
  \city{Hempstead, NY}
  \country{USA}}

\author{Di Liu}
\affiliation{%
  \institution{Norwegian University of Science and Technology}
  \city{Trondheim}
  \country{Norway}
}

\author{Amin Saremi}
\affiliation{%
  \institution{Ume\aa\ University}
  \city{Ume\aa}
  \country{Sweden}
}

\begin{abstract}
We consider latency attacks on object detectors, where the attacker's goal is not to corrupt a prediction but to make the system fail to respond in time, targeting real-time applications such as autonomous driving. Modern object detectors eliminate Non-Maximum Suppression (NMS) through one-to-one assignment or set prediction, removing the classical detector-side latency bottleneck exploited by prior latency (``sponge'') attacks. We show that NMS-free does not mean latency-robust: this architectural change does not eliminate the attack surface but relocates it downstream to multi-object tracking, whose data-association cost remains content dependent. We present \emph{TrackFlood}, a unified white-box overload attack against NMS-free detect-then-track pipelines spanning both one-to-one detectors (YOLOv10 and YOLO26) and query-based detectors (RT-DETR). TrackFlood recovers differentiable confidence tensors and optimizes perturbations that flood the tracker with spatially distributed phantom detections while leaving detector inference unchanged. Evaluated entirely on an NVIDIA Jetson AGX Orin (TensorRT FP16), detector latency remains essentially constant, whereas tracker latency increases substantially. At a standard imperceptible budget ($L_\infty{=}8/255$), a universal perturbation produces clearly measurable tracker overload but only modest end-to-end slowdown, without deadline misses for the association-dominated trackers; a separate higher-budget stress test drives severe end-to-end slowdowns and sustained deadline misses. We further evaluate a lightweight, architecture-agnostic bounded-admission layer that caps the tracker workload and largely restores end-to-end latency, at a non-trivial cost in admitted clean detections. Our results demonstrate that evaluating NMS-free perception systems requires considering downstream tracking and end-to-end timing, not detector inference alone.
\end{abstract}


\maketitle

\section{Introduction}
Adversarial machine learning has historically emphasized \emph{integrity}:
perturbations that cause a model to produce a wrong output. A complementary and
increasingly consequential axis is \emph{availability}, in which the attacker's
goal is not to corrupt a prediction but to make the system fail to respond in
time~\cite{shumailov2021sponge,shapira2023}; see~\cite{gu2026survey} for a
cross-domain survey of such availability threats and~\cite{meftah2025energy} for
a taxonomy of energy-latency attacks. Such latency (or ``sponge'',
energy-latency, slowdown) attacks now span sparsity-exploiting and input-adaptive
networks~\cite{krithivasan2020sparsity,haque2020ilfo,hong2021panda}, efficient
vision transformers~\cite{navaneet2024slowformer,yehezkel2024desparsify},
autoregressive language and vision-language
models~\cite{chen2022nmtsloth,chen2022nicgslowdown,gao2024verbose}, and LiDAR
and cooperative perception~\cite{liu2023slowlidar,wang2025cpfreezer}. They are
especially dangerous for perception
systems on embedded platforms in autonomous vehicles, where a missed
per-frame deadline can be safety-critical~\cite{ma2024slowtrack}.

For camera-based detection, the canonical latency vulnerability is Non-Maximum
Suppression (NMS), whose cost grows super-linearly with the number of candidate
boxes. Daedalus~\cite{wang2022daedalus}, Phantom Sponges~\cite{shapira2023}, and Overload~\cite{chen2023overload}
weaponize this with perturbations that add phantom boxes to overload NMS while
preserving the original detections, reporting up to a tenfold inference-time
increase; SlowTrack~\cite{ma2024slowtrack} and physical, projector-based
variants~\cite{muller2025detstorm} extend the threat to detect-then-track pipelines.
A recent architectural shift threatens to close this attack surface entirely.
Modern detectors increasingly \emph{remove} NMS: YOLOv10~\cite{wang2024yolov10}
and YOLO26~\cite{jocher2026yolo26} replace it with a learned one-to-one head, and DETR-style
models~\cite{carion2020detr,zhao2024rtdetr} use bipartite (Hungarian) matching and a
lightweight decoder. Both eliminate the quadratic post-processing blowup that
Phantom Sponges and Overload exploit, and both have a forward pass whose
compute is fixed regardless of how many objects appear. It is therefore natural
to ask whether NMS-free detection is inherently immune to latency attacks.

Our answer is no. We hypothesize that removing NMS does not eliminate the threat; it relocates
it, and provide, to our knowledge, the first end-to-end empirical demonstration \emph{for
NMS-free detectors} that the relocated bottleneck is the \emph{downstream multi-object
tracker} (SlowTrack~\cite{ma2024slowtrack} previously attacked trackers behind NMS-based
detectors).
The tracker's data association (gating, IoU/appearance matching, and Kalman
update) still scales with the number of detections it receives per frame. An
attacker who can no longer slow the detector can still flood it with phantom
detections that the fixed-FLOP forward pass happily emits, and thereby overload
the tracker; we call this attack \emph{TrackFlood}. From a real-time-systems perspective, the question is whether an
adversary can inflate the worst-case execution time of a pipeline stage through
its input content---and whether classical admission control can restore a bound.
We make the following contributions in this paper:
\begin{itemize}[nosep] 
\item We formulate TrackFlood, a unified white-box overload attack that works across two
distinct NMS-free detector families: anchor-based one-to-one heads
(YOLOv10, YOLO26) and query-based set prediction (RT-DETR). The core technical
step is recovering each detector's raw, gradient-bearing confidence tensor,
which the standard inference API hides behind detached features and top-$k$
selection.
\item We demonstrate end-to-end that TrackFlood relocates latency from the detector to the downstream tracker. Under the deployment-realistic universal attack at an $L_\infty=8/255$ budget, measured on-device, the attack substantially inflates tracker latency while end-to-end slowdown stays modest and the association-dominated trackers miss no deadlines; severe end-to-end slowdowns and sustained deadline misses appear only at the higher-budget stress test.
\item We show the overload magnitude is \emph{regime-dependent}, governed by the
achievable detection flood (attack strength $\times$ scene density $\times$
threshold) rather than the head type: RT-DETR's bounded query set makes it
hardest to flood on sparse scenes yet its high output density makes it the most
overloaded on dense scenes. This is, to our knowledge, the first comparative
overload evaluation of NMS-free detectors, reported with per-sequence variance.
\item We measure the attack entirely on-device (Jetson AGX Orin, TensorRT FP16),
showing the GPU detector latency is unchanged while the downstream tracker is the
attacked stage, and we evaluate a bounded-admission layer as a defense on the
same platform.
\end{itemize}

\section{Background and Threat Model}
\subsection{NMS-free detectors}
\textbf{One-to-one heads (YOLOv10, YOLO26).} These detectors retain a dense
anchor grid (e.g., 8400 anchors at $640\times640$) but train a ``one-to-one''
head with a learned assignment so that, at inference, duplicate suppression is
implicit and NMS is dropped. The forward cost is fixed by the network, not by
the scene.

\textbf{Query-based set prediction (RT-DETR).} A transformer decoder emits a
fixed set of $Q$ predictions (here $Q=300$), each a box and class distribution.
Bipartite matching at training time enforces one prediction per object,
removing NMS. The number of prediction slots is a hard architectural cap.

This architectural change also changes where latency attacks can take effect. In NMS-based detectors, producing many candidate boxes directly increases detector latency because the detector must run content-dependent NMS. In contrast, NMS-free detectors evaluate a fixed set of prediction slots each frame, so increasing the number of boxes above threshold changes the outputs but not the detector's main computation. The attack surface therefore shifts downstream to the multi-object tracker, whose cost still depends on the number and spatial spread of detections it receives. Accordingly, our method does not slow a pure standalone NMS-free detector itself; it overloads the tracker in a detect-then-track pipeline while detector inference remains essentially unchanged.

\subsection{Detect-then-track pipeline and tracker cost structure}
\label{sec:bg-tracker}
A typical autonomous-driving perception stack feeds detector outputs into a
multi-object tracker (MOT) that maintains a persistent identity for each object
across frames. Modern real-time trackers follow the \emph{tracking-by-detection}
paradigm~\cite{bewley2016sort}: every frame, the tracker (i) predicts each existing track's new
position with a per-track \emph{Kalman filter}~\cite{kalman1960} (a constant-velocity motion
model), (ii) \emph{associates} the incoming detections to those predicted tracks,
and (iii) updates matched tracks, spawns tracks for unmatched detections, and
retires tracks unseen for a buffer of frames. The association step---the focus of
this paper---solves a bipartite matching (Hungarian assignment~\cite{kuhn1955hungarian}) over a cost
matrix built from every (track, detection) pair. Constructing this matrix is
$O(TD)$ in the number of tracks $T$ and detections $D$, and Hungarian-style
assignment is worst-case cubic in the larger problem dimension. Increasing $D$
therefore raises current-frame association work directly, while persistent false
tracks can increase $T$ in subsequent frames, producing \textbf{cumulative and
potentially super-linear growth}. Unlike the NMS-free detector, whose forward
pass is fixed-structure, the tracker is therefore \emph{content-dependent}:
flooding it with detections directly increases its per-frame workload.

We evaluate four representative tracker families that differ in how much of their
runtime is detection-dependent versus detection-independent: association-dominated
(\emph{ByteTrack}~\cite{bytetrack}, targeted by SlowTrack~\cite{ma2024slowtrack},
and \emph{OC-SORT}~\cite{ocsort}), appearance-assisted (\emph{DeepOCSORT}
~\cite{deepocsort}, which runs a ReID embedding per detection, cost $\propto D$),
and motion-compensated (\emph{BOT-SORT}~\cite{botsort}, which adds Global Motion
Compensation---an optical-flow stage that operates on the \emph{image}, so its
large cost is \emph{independent of $D$} and dilutes the relative multiplier even
while association is overloaded). This spread lets us isolate where overload
arises, and is why timing is not portable across trackers
(Section~\ref{sec:tracker2}).

\begin{figure*}[!htbp]
\centering
\includegraphics[width=\textwidth]{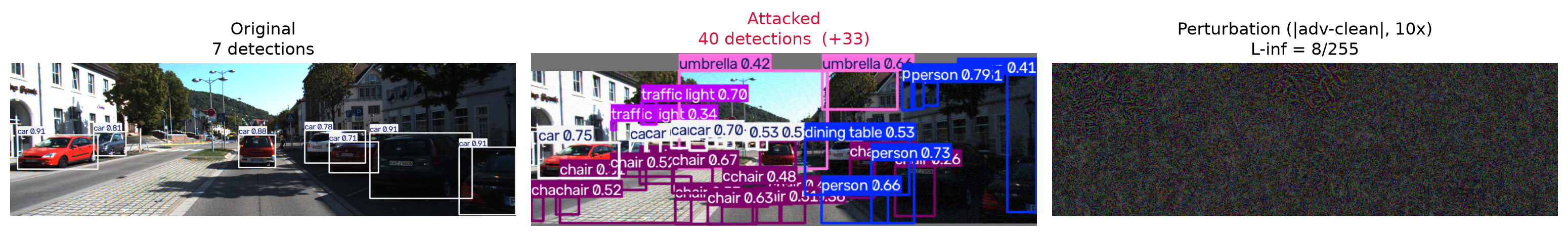}
\caption{YOLO26 on real KITTI sequence 0011 under an invisible digital $L_\infty=8/255$ perturbation confined to the scene content.
Left: original detections (7). Center: attacked (40). Right: the $10\times$-amplified perturbation. Detector inference latency is unchanged; the flood of phantom detections inflates only the downstream tracker latency.}
\label{fig:y26}
\end{figure*}
\subsection{Threat model}
\label{sec:threat}
We assume a white-box attacker who can add a digital perturbation
$\delta$ to the camera frame within an $L_\infty$ budget $\epsilon=8/255$.
We adopt $L_\infty=8/255$ as a \emph{standard bounded digital-perturbation
budget} in adversarial vision: it caps the change of each 8-bit pixel channel at
eight intensity levels ($\approx0.031$ on normalized $[0,1]$ inputs). We use the \emph{same} budget for the
optimized attack and the random-noise control, so the comparison is at matched
perturbation strength; the noise check itself is reported in \S\ref{sec:eps}. The $L_\infty$ max-norm is a per-pixel guarantee and is
not inherited from SlowTrack, which instead reports an average $L_2$ norm
($\approx0.021$) for imperceptibility~\cite{ma2024slowtrack}; the two metrics are
not directly comparable, and we choose $L_\infty$ because it gives a
pixel-wise worst-case bound that is the convention in adversarial
robustness~\cite{madry2018pgd} and in the sponge/latency-attack
literature~\cite{shapira2023,chen2023overload}. Section~\ref{sec:eps} reports a
sensitivity study over $\epsilon\in\{4,8,16,32\}/255$ to confirm the relocation
effect persists across perturbation budgets. Crucially, the perturbation and the
attack objective are restricted to the \emph{scene-content region} of the
letterboxed input, excluding the grey padding that a real camera attacker cannot
control. The attacker's goal is availability: maximize downstream tracker
latency, not cause misdetection. The detector weights are frozen; only pixels
are optimized. Figure~\ref{fig:y26} shows one example: under the $L_\infty{=}8/255$ attack the detection count rises from 7 to 40, and the right panel visualizes the perturbation itself, amplified $10\times$ (i.e. $10\times|\text{adv}-\text{clean}|$) for visualization, since at true scale it is imperceptible to the human eye.

\section{Method}
\label{sec:method}
TrackFlood has three components: recovering differentiable per-slot confidences
from each NMS-free detector family (\S\ref{sec:recover}), an overload objective
that floods the tracker with spatially spread phantom detections
(\S\ref{sec:objective}), and a universal, image-agnostic variant that is trained
offline and applied unchanged to live frames (\S\ref{sec:universal}).
\subsection{Recovering differentiable confidence tensors}
\label{sec:recover}
Our optimization requires differentiable per-slot detection
confidences, which standard inference APIs do not expose because they
detach intermediate features or return only post-processed outputs. We
therefore recover the raw confidence tensor before post-processing,
obtaining differentiable confidence scores for all prediction slots
(8400 anchors for one-to-one detectors (YOLO variants) or 300 queries for RT-DETR),
which serve as the input to the attack objective.

\textbf{Optimization and measurement backends.} Attack optimization runs in
PyTorch, differentiating through the raw per-slot confidences recovered above;
deployment timing runs on TensorRT FP16 engines built from the same weights, and
all detections-per-frame statistics are counted on the engine outputs. For the
one-to-one heads we recover the per-anchor scores of the one-to-one branch
\emph{before} top-$k$ selection, using a forward pre-hook that re-runs the head
on non-detached feature maps; for RT-DETR we read the last decoder layer's
sigmoid class scores directly. The detector's $1.00\times$ latency ratio is
structural---fixed input resolution and a fixed slot count (8400 anchors or 300
queries) make the forward pass data-independent---and the measured detector
ratio stays within $0.99$--$1.01$ across all detectors, budgets, and captures.

\subsection{Overload objective}
\label{sec:objective}
Let $b_i$ be the box center of slot $i$ (in normalized image coordinates),
$\tau$ the detection threshold, and $m_i\in\{0,1\}$ a content mask (Section~\ref{sec:threat}). We define a soft detection indicator
$w_i = \sigma\!\big(\alpha(c_i-\tau)\big)\, m_i$ and two loss terms:
\begin{equation}
\mathcal{L}_{\text{flood}} = \sum_i w_i, \qquad
\mathcal{L}_{\text{spread}} = \mathrm{Var}_w(b_x) + \mathrm{Var}_w(b_y),
\end{equation}
where $\mathrm{Var}_w$ is the $w$-weighted spatial variance of box centers. The \emph{flood} term (a differentiable proxy for the above-threshold detection count, since a hard count $c_i{>}\tau$ is non-differentiable) maximizes how many
detections exist, while the \emph{spread} term rewards spatially separated
detections that resolve into distinct tracks rather than collapsing into one---the
same count but more association and Kalman-update work per frame. The attack minimizes
$-(\mathcal{L}_{\text{flood}} + \lambda\,\mathcal{L}_{\text{spread}})$ via
projected gradient descent (PGD)~\cite{madry2018pgd} under the $L_\infty$ budget. For query-based RT-DETR, whose $300$ bipartite-matched queries resist the pure
count objective, we use a per-slot hinge variant that drives every query's
confidence toward $\tau+\gamma$:
\begin{equation}
\mathcal{L}_{\text{margin}} = \sum_i \mathrm{ReLU}\!\big((\tau+\gamma) - c_i\big)\, m_i
- \lambda\,\mathcal{L}_{\text{spread}}.
\end{equation}

\begin{tcolorbox}[colback=gray!6,colframe=gray!45,boxrule=0.4pt,
arc=1.5pt,left=4pt,right=4pt,top=3pt,bottom=3pt,
fonttitle=\bfseries\small,title=Intuition: why flood \emph{and} spread]
\small
With $\tau{=}0.25,\alpha{=}50$, the soft indicator $w_i$ reads $\approx0$ below
threshold and $\approx1$ above, giving a differentiable proxy for the detection
count that a hard count ($c_i{>}\tau$) cannot provide:
\begin{center}
\begin{tabular}{ccccc}
\toprule
$c_i$ & 0.05 & 0.24 & 0.26 & 0.90 \\
$w_i$ & 0.00 & 0.38 & 0.62 & 1.00 \\
\bottomrule
\end{tabular}
\end{center}
\textbf{Flood} sums these (pushing $c=[.05,.24,.26,.90]\to[.30,.40,.50,.92]$
raises $\mathcal{L}_{\text{flood}}$ from $2.0$ to $3.9$). \textbf{Spread} rewards
placing them apart: four boxes at $x{=}[48,50,51,52]$ collapse into one cheap
track, whereas $x{=}[10,40,60,90]$ force four distinct tracks---the same count
but $4\times$ the association work.
\end{tcolorbox}

\textbf{PGD update}. At each step we take $\delta \leftarrow \mathrm{clip}_{[-\epsilon,\epsilon]}
\big(\delta - s\cdot\mathrm{sign}(\nabla_\delta \mathcal{L})\big)$, mask $\delta$
to the content region, and re-project to a valid image. We use
$\alpha{=}50$, $s{=}2/255$, $\lambda{=}0.1$, $\gamma{=}0.3$, and 40 (YOLO) or
60 (RT-DETR) iterations.

\subsection{Universal (image-agnostic) attack}
\label{sec:universal}
The per-frame attack solves a fresh $\delta$ for each input. A more operationally
realistic but generally weaker transfer setting, following Phantom
Sponges~\cite{shapira2023}, computes a universal perturbation~\cite{moosavi2017uap}
offline and applies it unchanged to every live frame, with no per-frame
optimization. We train a \emph{separate} universal perturbation for each detector
by averaging that detector's overload gradients over mini-batches of training
frames: at each step we take one $L_\infty$ sign step and project $\delta$ into
the $\epsilon$ ball and the (shared) scene-content region. Each detector's
$\delta$ is then frozen and evaluated on a disjoint held-out split, so the
reported effect is pure transfer to unseen frames.

\section{Experiments}

\subsection{Setup}
We evaluate three detectors---YOLOv10-n, YOLO26-n (one-to-one) and RT-DETR-l
(query-based)---on real KITTI~\cite{geiger2012kitti} driving sequences (17-sequence training split, 4
disjoint held-out test sequences for the universal attack), feeding detections
to ByteTrack~\cite{bytetrack} (the tracker used by SlowTrack~\cite{ma2024slowtrack}) at $\tau{=}0.25$ unless noted.
We report mean detections/frame, median detector and tracker \texttt{update()}
latency, mean active tracks, and the tracker-latency multiplier
(adversarial/clean); timing the tracker in isolation avoids drowning its
few-millisecond signal in the detector's larger latency. All timing data and all detection counts in the tables are collected on-device (Jetson AGX Orin, TensorRT FP16); the only exception is the small noise check in \S\ref{sec:eps}.
Reported multipliers are \emph{ratios of medians} (attacked over clean on
identical frames). End-to-end latency is the sum of the CUDA-event-timed
detector call and the host-timed tracker \texttt{update()}, excluding image
preprocessing and output decoding; the detector is additionally benchmarked with
a 300-repetition fixed-input microbenchmark, and every number follows a
20-iteration warmup with clocks locked. Trackers use their stock Ultralytics
configurations (\texttt{track\_buffer}${=}30$); the ByteTrack rows in
Tables~\ref{tab:eps_jetson_full} and~\ref{tab:univ}(a) use our vendored
ByteTrack-lineage build, while panel~(b) of Table~\ref{tab:univ} uses the
Ultralytics ByteTrack build.

\subsection{Central result: regime-dependent overload}
\label{sec:regime}
Our central finding is that removing Non-Maximum Suppression (NMS) relocates
the latency attack surface downstream to the tracker for every NMS-free
detector we study. The resulting overload is \emph{regime-dependent}: its
magnitude is governed not by detector head type alone, but by the achievable
detection flood. The achievable detection flood, in turn, depends on (i) attack
strength (per-frame $>$ universal), (ii) scene density and detector output density
(dense multi-object scenes $>$ sparse scenes), and (iii) the detection
threshold.

Consequently, no detector is uniformly the most or least vulnerable.
RT-DETR's bounded query set makes it comparatively resistant to flooding under
sparse scenes and the weaker universal attack, yet its higher output density produces
the largest tracker overload in dense scenes, reaching a dense-pool aggregate
tracker-latency multiplier of \textbf{2.2$\times$}
(Table~\ref{tab:univ}). Thus, the observed slowdown is determined primarily by
the achievable detection flood rather than detector head type itself. This observation provides a unifying explanation for the behavior of all three NMS-free detector families studied. Detector architecture determines how easily the output can be flooded, but once a sufficient detection flood is achieved, overload is governed primarily by the downstream tracker rather than by the detector itself. Consequently, detector-side latency immunity does not imply end-to-end latency robustness. The remaining subsections examine each of these operating regimes on the Jetson AGX
Orin.

\subsection{Perturbation-budget sensitivity}
\label{sec:eps}

We sweep $\epsilon\in\{4,8,16,32\}/255$ on the Jetson AGX Orin (TensorRT FP16) for all three detectors and four tracker families (five configurations, counting BOT-SORT with GMC on/off; Table~\ref{tab:eps_jetson_full}). The \emph{detector} runs on the \emph{GPU} with clocks locked
(\texttt{nvpmodel -m 0}, \texttt{jetson\_clocks}) and warm-up; its GPU latency is
measured with device-side CUDA events and explicit synchronization over many
repetitions, following NVIDIA's recommended timing procedure~\cite{geifman2020timing}. The \emph{tracker} is a data-association
stage that runs on the Orin's on-board \emph{ARM CPU}; the end-to-end value is
the sum of the two. It is this data association cost that the attack
inflates while the GPU detector cost stays fixed. 

Three findings emerge from the on-device measurements. First, the detector latency ratio remains within $0.99$--$1.01\times$ at every perturbation budget for all three architectures (the structural fixed-resolution, fixed-slot-count property of \S\ref{sec:method}): the perturbation changes only how many detections exceed the threshold $\tau$, not the detector's fixed-FLOP computation, so the overload is entirely downstream in the tracker. Second, at the 8/255 security budget, the ByteTrack end-to-end multiplier reaches 1.47$\times$ (YOLOv10-n), 1.18$\times$ (YOLO26-n), and 1.09$\times$ (RT-DETR-l), with zero missed deadlines, demonstrating measurable but modest overload. Third, under the higher-budget stress test (32/255), the attack begins to violate the 33 ms (30 FPS) real-time deadline: ByteTrack reaches 6.90$\times$ (YOLOv10-n, 62\% missed deadlines), while the association-heavy OC-SORT and DeepOCSORT reach 13.54$\times$ and 13.15$\times$, respectively, with 100\% deadline misses—the largest slowdowns observed in this paper. As shown by the BOT-SORT rows in Table~\ref{tab:eps_jetson_full}, enabling GMC compresses the observed relative latency multiplier because the fixed optical-flow stage dominates runtime, masking the contribution of association latency while producing the highest absolute p99 latency of any tracker at the $4/255$ and $8/255$ budgets in Table~\ref{tab:eps_jetson_full} (e.g., $43$\,ms at $8/255$ for YOLOv10-n; at $32/255$ OC-SORT and DeepOCSORT exceed it). The security-budget and stress-test regimes are therefore distinct: overload is clearly measurable at the perturbation budget of 8/255, whereas catastrophic deadline failures occur only under the higher-budget stress test. Because BOT-SORT with GMC already incurs deadline misses at low perturbation budgets due to its large detection-independent GMC stage, we interpret attack pressure through the increase relative to the low-budget baseline rather than the absolute miss fraction alone. The YOLO26 rows saturate between $16/255$ and $32/255$ (e.g., ByteTrack $3.08\times\to3.10\times$) because the achievable flood itself saturates: its universal delta reaches $88.3$ detections/frame at $16/255$ and only $98.2$ at $32/255$, whereas YOLOv10's flood continues to grow ($65.1\to238.1$), so its multipliers keep rising. A small same-budget noise check (uniform and Rademacher at $8/255$, one draw each, YOLOv10-n on six held-out frames of sequence 0011, counted with the PyTorch FP32 model on a CPU host rather than on the TensorRT engine) leaves the detection count essentially unchanged ($7.5$ vs.\ $7.3$/$7.7$ detections per frame for clean vs.\ noise), whereas the optimized perturbation floods ($10.5$). This is consistent with the flood being attributable to the adversarial optimization rather than the perturbation magnitude, but at this sample size it is indicative only; we did not repeat the on-device timing under noise.

\begin{table*}[htbp]
\centering
\begin{small}
\caption{Perturbation-budget sweep across tracker families (\emph{universal}
attack). Each cell: median end-to-end multiplier vs.\ clean (p99 ms /
deadline-miss fraction; $33$\,ms budget), $\epsilon\in\{4,8,16,32\}/255$,
$\tau=0.25$. $8/255$ is the security operating point; $32/255$ is a stress test.}
\label{tab:eps_jetson_full}
\setlength{\tabcolsep}{4pt}
\begin{tabular}{@{}llcccc@{}}
\toprule
Detector & Tracker & $4/255$ & $8/255$ & $16/255$ & $32/255$ \\
\midrule
\multirow{5}{*}{\shortstack[l]{YOLO\\v10-n}}
  & ByteTrack          & 1.02$\times$ (6/0.00)  & \textbf{1.47$\times$} (10/0.00) & 2.59$\times$ (16/0.00) & 6.90$\times$ (40/0.62) \\
  & OC-SORT            & 1.00$\times$ (12/0.00) & 2.11$\times$ (26/0.00) & 4.37$\times$ (43/0.17) & 13.54$\times$ (109/1.00) \\
  & DeepOCSORT         & 1.00$\times$ (13/0.00) & 2.12$\times$ (25/0.00) & 4.41$\times$ (41/0.30) & 13.15$\times$ (112/1.00) \\
  & BOT-SORT (GMC on)  & 1.01$\times$ (40/0.07) & 1.14$\times$ (43/0.12) & 1.51$\times$ (53/0.70) & 3.30$\times$ (103/1.00) \\
  & BOT-SORT (GMC off) & 1.02$\times$ (7/0.00)  & 1.57$\times$ (11/0.00) & 2.95$\times$ (19/0.00) & 8.21$\times$ (50/0.95) \\
\midrule
\multirow{5}{*}{YOLO26-n}
  & ByteTrack          & 1.01$\times$ (6/0.00)  & \textbf{1.18$\times$} (7/0.00)  & 3.08$\times$ (19/0.00) & 3.10$\times$ (19/0.00) \\
  & OC-SORT            & 1.01$\times$ (13/0.00) & 1.37$\times$ (16/0.00) & 5.17$\times$ (47/0.69) & 5.09$\times$ (43/0.72) \\
  & DeepOCSORT         & 1.01$\times$ (14/0.00) & 1.37$\times$ (17/0.00) & 5.26$\times$ (47/0.77) & 5.18$\times$ (47/0.88) \\
  & BOT-SORT (GMC on)  & 1.01$\times$ (40/0.06) & 1.09$\times$ (42/0.09) & 1.77$\times$ (61/0.95) & 2.18$\times$ (68/1.00) \\
  & BOT-SORT (GMC off) & 1.01$\times$ (7/0.00)  & 1.24$\times$ (8/0.00)  & 3.63$\times$ (23/0.00) & 3.67$\times$ (23/0.00) \\
\midrule
\multirow{5}{*}{RT-DETR-l}
  & ByteTrack          & 1.00$\times$ (19/0.00) & \textbf{1.09$\times$} (21/0.00) & 1.41$\times$ (31/0.00) & 1.90$\times$ (38/0.26) \\
  & OC-SORT            & 1.01$\times$ (21/0.00) & 1.11$\times$ (23/0.00) & 1.51$\times$ (38/0.14) & 2.19$\times$ (48/0.78) \\
  & DeepOCSORT         & 1.00$\times$ (22/0.00) & 1.10$\times$ (23/0.00) & 1.44$\times$ (37/0.13) & 2.11$\times$ (49/0.70) \\
  & BOT-SORT (GMC on)  & 1.01$\times$ (52/0.70) & 1.06$\times$ (53/0.85) & 1.26$\times$ (65/1.00) & 1.76$\times$ (79/1.00) \\
  & BOT-SORT (GMC off) & 1.01$\times$ (20/0.00) & 1.09$\times$ (21/0.00) & 1.39$\times$ (31/0.00) & 1.94$\times$ (40/0.31) \\
\bottomrule
\end{tabular}
\end{small}
\end{table*}

\subsection{Universal attack on real KITTI sequences}
For each detector, we train a separate universal perturbation on the same
17-sequence training pool and apply it, frozen, to 593 held-out frames at
$\tau{=}0.25$ (Table~\ref{tab:univ}). A detector-specific fixed pattern
successfully floods each of the three NMS-free detectors on unseen frames,
yielding aggregate tracker-latency multipliers of $1.25\times$ for YOLO26,
$1.59\times$ for YOLOv10, and $2.22\times$ for RT-DETR. These effects are weaker
than those of the image-specific attack, reflecting the usual trade-off between
cross-frame generalization and attack strength. In Table~\ref{tab:univ},
``Tracks'' reports the mean active-track count after \texttt{update()}; it is
lower than the detection count because not every admitted detection immediately
becomes a persistent track.

\begin{table}[htbp]
\centering
\begin{small}
\caption{Universal overload on \emph{real KITTI} at $L_\infty{=}8/255$,
$\tau{=}0.25$; each row uses that detector's own frozen universal perturbation.
\textbf{(a)} Aggregate over the 593-frame held-out pool (c$\to$a: clean vs.\
attack). \textbf{(b)} Per-clip: tracker-latency and end-to-end multipliers;
slowdown grows with density (largest on clip~0020).}
\label{tab:univ}
\setlength{\tabcolsep}{2.5pt}
\begin{tabular}{@{}lccccc@{}}
\multicolumn{6}{l}{\emph{(a) Dense held-out aggregate, 593-frame pool}}\\
\toprule
           & Det./fr.  &  Tracks   & Trk.\ ms & Trk. & e2e \\
Detector & c$\to$a & c$\to$a & c$\to$a & mult. & mult. \\
\midrule
YOLO26-n  & $6.0 \to 9.1$   & $2.5 \to 6.1$  & $2.35 \to 2.93$ & $1.25\times$ & $1.11\times$ \\
YOLOv10-n & $5.9 \to 14.2$  & $2.3 \to 9.1$  & $2.26 \to 3.60$ & $1.59\times$ & $1.24\times$ \\
RT-DETR-l & $22.8 \to 65.4$ & $7.8 \to 28.8$ & $3.06 \to 6.80$ & $2.22\times$ & $1.22\times$ \\
\bottomrule
\end{tabular}
\vspace{2pt}

\setlength{\tabcolsep}{3.2pt}
\begin{tabular}{llccccc}
\multicolumn{7}{l}{\emph{(b) Per held-out clip}}\\
\toprule
Detector & Metric & 0011 & 0005 & 0013 & 0020 & mean$\pm$std \\
\midrule
YOLO26-n  & tracker & $1.59$ & $2.05$ & $1.28$ & $1.94$ & $\mathbf{1.72{\pm}0.30}$ \\
          & e2e     & $1.12$ & $1.18$ & $1.06$ & $1.19$ & $1.14{\pm}0.05$ \\
\addlinespace[1pt]
YOLOv10-n & tracker & $2.05$ & $2.73$ & $1.69$ & $3.54$ & $\mathbf{2.50{\pm}0.70}$ \\
          & e2e     & $1.22$ & $1.30$ & $1.13$ & $1.52$ & $1.29{\pm}0.14$ \\
\addlinespace[1pt]
RT-DETR-l & tracker & $1.33$ & $1.62$ & $1.06$ & $2.28$ & $\mathbf{1.57{\pm}0.45}$ \\
          & e2e     & $1.03$ & $1.05$ & $1.01$ & $1.10$ & $1.05{\pm}0.03$ \\
\bottomrule
\end{tabular}
\vspace{1pt}
\end{small}
\end{table}

\subsection{Operating detection density governs susceptibility}
\label{sec:density}
\label{sec:recall}
We define \emph{operating output density} as the number of detections per frame
at the configured threshold; unlike recall, this quantity does not require
matching predictions to ground truth. On cluttered KITTI scenes RT-DETR has
substantially higher output density than the nano YOLO heads ($\sim$$25$ vs.\
$\sim$$6$ detections/frame at $\tau{=}0.25$), and this larger clean pool is what
the attack amplifies---RT-DETR floods from $22.8$ to $65.4$ detections per
frame, the largest of any detector, giving the highest tracker multiplier
($2.22\times$). This is a regime-dependent property: Table~\ref{tab:univ}(b) shows per-clip tracker slowdown for every
detector ($1.72{\pm}0.30$ YOLO26, $2.50{\pm}0.70$ YOLOv10, $1.57{\pm}0.45$
RT-DETR) growing with density---largest on the crowded clip~0020
($1.94$--$3.54\times$), smallest on sparse clip~0013 ($1.06$--$1.69\times$). Because the fixed-cost GPU detector accounts for a large share of the per-frame time (most of it for RT-DETR-l, roughly half for the nano YOLO heads), even substantial tracker slowdowns translate into relatively modest ByteTrack end-to-end slowdowns ($1.0$--$1.3\times$), which is why the attacked \emph{stage} must be reported. This resolves the apparent conflict with reports that DETRs do
not expose a \emph{detector-side} latency surface (the DETR analysis in~\cite{underload}): our payload is the tracker, and RT-DETR's fixed-cost forward feeds it the heaviest load in the high-flood regime.

\emph{Experimental regimes and measurement batches.} The universal attack uses real KITTI (universal deltas trained on a pooled training split, evaluated on a disjoint held-out pool). All tables share one timing protocol but aggregate \emph{different captures and frozen-delta realizations}, so multipliers are not directly comparable across tables. Tables~\ref{tab:eps_jetson_full}, \ref{tab:univ}(a), and \ref{tab:defense_jetson} measure the same 593-frame dense held-out pool, the held-out 20\% of a single 80/20 split over pooled real KITTI tracking frames; the split was assembled once and every pooled capture reuses those exact frames, and the original universal deltas are trained only on the 80\% training half. We separate \emph{which frames were evaluated} from \emph{how they were selected}: the full per-frame manifest of the pool (sequence and frame identifiers for all 593 frames) accompanies our artifact.
Table~\ref{tab:eps_jetson_full} sweeps budgets with a separately trained universal delta per budget (at $8/255$, RT-DETR floods $22.8\to36.8$ detections/frame), Table~\ref{tab:defense_jetson} reuses that realization, and Table~\ref{tab:univ}(a) uses the original single universal delta per detector (RT-DETR floods $22.8\to65.4$). Table~\ref{tab:univ}(b) is a separate per-clip capture over the four \emph{full} held-out sequences ($373+297+340+837=1{,}847$ frames) with its own delta realization. Within each capture, clean and attacked runs share identical frames, engine, and session, and the end-to-end multiplier is always \emph{below} the tracker-stage multiplier because detector time is unchanged; across captures the same nominal setting can differ (the RT-DETR-l + ByteTrack $8/255$ end-to-end multiplier reads $1.09\times$ in Table~\ref{tab:eps_jetson_full}, $1.22\times$ in Table~\ref{tab:univ}(a), $1.05\times$ in Table~\ref{tab:univ}(b), and $1.08\times$ in Table~\ref{tab:defense_jetson}), reflecting different flood realizations rather than contradiction. For the same reason the pooled YOLO multipliers in panel~(a) can fall below every per-clip value in panel~(b): the deployment-capture delta floods less on the dense pool (YOLO26 clean $6.0$, flooded to $9.1$) than the per-clip realization does on the full sequences (clean $3.5$--$6.4$, flooded to $8.0$--$21.9$ across clips). The pool is genuinely denser than the full held-out sequences ($6.0$ vs.\ $5.4$ frame-weighted clean detections/frame for YOLO26, and $22.8$ vs.\ $16.9$ for RT-DETR-l).

\subsection{Generalization across tracker families}
\label{sec:tracker2}
To determine whether the observed overload is a property of the downstream
\emph{association structure} rather than of a particular tracker implementation,
we evaluate four representative tracker families
(Table~\ref{tab:tracker}).

Association-dominated trackers, including ByteTrack and OC-SORT, exhibit
slowdowns whenever the attack successfully increases the detection count.
At the $8/255$ security budget (the same capture as Table~\ref{tab:eps_jetson_full}),
OC-SORT reaches a tracker-latency multiplier of $2.91\times$ and ByteTrack
$2.04\times$, closely tracking the increase in association work. DeepOCSORT
exhibits a similar relative slowdown ($2.86\times$) with a modestly higher
absolute cost: its median \texttt{update()} latency is $4.23$\,ms clean and
$12.08$\,ms under attack, versus $3.81$ and $11.10$\,ms for OC-SORT.
We attribute the difference to its per-detection appearance-ReID stage, but the gap (about $0.4$--$1$\,ms) is small for a per-detection embedding network and we did not profile that stage separately; because the gap is small, the two trackers'
end-to-end p99 values nearly coincide in Table~\ref{tab:eps_jetson_full}.

BOT-SORT performs Global Motion Compensation (GMC) before association. We isolate
its effect with a \emph{paired} run on four held-out clips that genuinely flood
under the YOLOv10 universal delta ($3$--$6\to20$--$27$ det./frame), feeding the
\emph{same cached detections} to GMC-on and GMC-off ($60$ frames, $15$ reps, so
detection counts are identical across the two). With GMC \emph{off}, the attack
raises \texttt{update()} from $1.2$--$1.7$ to $3.7$--$4.9$\,ms
($\mathbf{3.08\times}$ mean, per-clip $2.16$--$3.81\times$, all CIs above $1$).
With GMC \emph{on}, the detection-independent optical-flow stage
($\sim$12--21\,ms) dominates, diluting the same flood to $\mathbf{1.23\times}$
($1.16$--$1.31\times$, CIs above $1$ but far smaller). The fixed GMC stage thus
masks the association overload without removing it---the dilution effect measured
with paired timing, free of the spurious sub-$1\times$ ratios of unpaired runs.

Overall, the attack impact scales with the fraction of runtime spent on
detection-dependent association. It is therefore largest for the
association-dominated trackers commonly deployed in real-time systems and is
diluted only when a large fixed, detection-independent stage dominates the
execution time, as in BOT-SORT with GMC.

\begin{table}[htbp]
\centering
\begin{small}
\caption{Universal attack across four tracker families ($\tau{=}0.25$;
YOLOv10; \texttt{update()} multiplier is attack/clean). The
ByteTrack/OC-SORT/DeepOCSORT rows are at $8/255$ on the 593-frame dense pool,
the same capture as Table~\ref{tab:eps_jetson_full}'s $8/255$ column. The
BOT-SORT rows are the paired GMC on/off comparison of \S\ref{sec:tracker2} on
four flooding clips.}
\label{tab:tracker}
\setlength{\tabcolsep}{3pt}
\begin{tabular}{@{}llc@{}}
\toprule
Tracker & Cost profile & Mult. \\
\midrule
ByteTrack          & IoU association + Kalman            & $2.04\times$ \\
OC-SORT            & Motion-only association & $2.91\times$ \\
DeepOCSORT         & Association + appearance ReID       & $2.86\times$ \\
BOT-SORT (GMC on)  & GMC + association      & $1.23\times$ \\
BOT-SORT (GMC off) & IoU association + Kalman             & $3.08\times$ \\
\bottomrule
\end{tabular}
\end{small}
\end{table}

\subsection{Qualitative results}
Fig.~\ref{fig:y26} illustrates the relocated attack surface: the clean frame
yields seven detections, while the perturbed frame---at the same detector
FLOPs and latency---emits 40, forcing the tracker to gate and associate a much
larger detection set. The bottleneck moves from detection to tracking.

\subsection{Cumulative overload and deadline pressure}
\label{sec:cumulative}
\begin{figure}[htbp]
\centering
\includegraphics[width=\columnwidth]{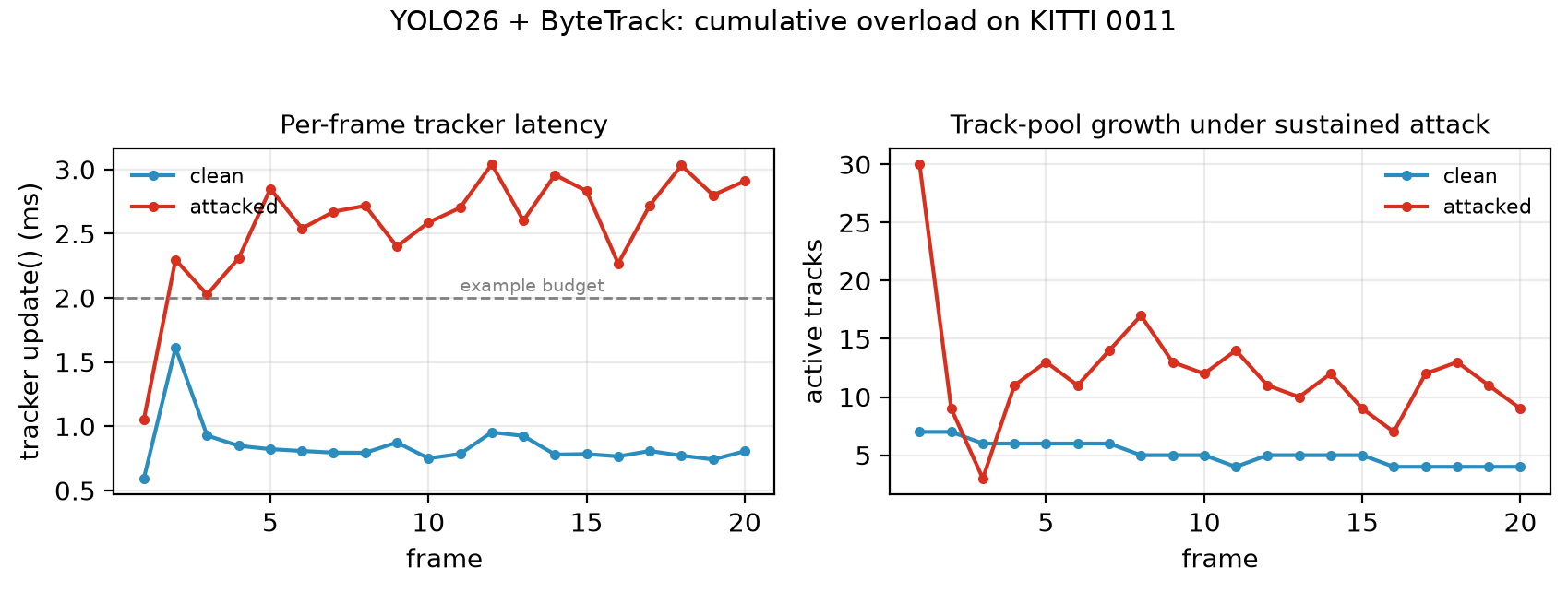}
\caption{Cumulative overload on KITTI 0011 (YOLO26 + ByteTrack). Left: per-frame
tracker \texttt{update()} latency stays elevated under attack (red) vs.\ clean
(blue). Right: the active track pool stays inflated for the whole sequence.}
\label{fig:deadline}
\end{figure}

A single frame's tracker cost is modest; the safety relevance lies in the
\emph{cumulative} effect. Under sustained attack, phantom tracks persist for the
track buffer, keeping the active pool inflated and latency elevated frame after
frame (Fig.~\ref{fig:deadline}, KITTI~0011/YOLO26: clean $\sim4$ tracks at
$\sim0.8$\,ms vs.\ 10--17 tracks at 2.5--3.0\,ms under attack). This converts a
benign per-frame cost into repeated deadline pressure, worst on dense scenes and
for the high-density RT-DETR head.

\subsection{On-device protocol (Jetson AGX Orin / TensorRT)}
\label{sec:jetson}
All timing is measured on-device on an NVIDIA Jetson AGX Orin (JetPack~6,
CUDA~12, TensorRT~10, MAXN + \texttt{jetson\_clocks}). Detectors run on the GPU
as TensorRT FP16 engines and are timed with device-side CUDA events; the tracker
\texttt{update()} runs on the Arm CPU and is timed with a single-core host clock;
end-to-end latency is their sum. The timed \texttt{update()} stage (the reported
``tracker latency'') covers all host-side work between receiving detections and
emitting tracks---confidence filtering, GMC, Kalman predict/update, ReID, and
Hungarian matching---not the assignment kernel in isolation. All numbers use a
20-iteration warmup and are medians over the reported frame set.

\section{Defenses}
\label{sec:def}

Two factors set how much a detector can be overloaded: the number of prediction
slots (YOLO's 8400 anchors vs.\ RT-DETR's 300 queries, so RT-DETR resists
flooding on \emph{sparse} scenes) and---dominant on realistic scenes---the
operating output density (\S\ref{sec:density}), which makes RT-DETR's stream the
heaviest tracker load. Bounded queries are thus not a reliable defense, and
NMS-centric mitigations (candidate caps~\cite{chen2023overload}, NMS-time limits, or constant-time post-processing proposed against NMS timing leakage~\cite{biton2023variable}) are irrelevant here;
defenses must instead bound the detections entering association. More generally, the defense objective differs fundamentally from detector-centric latency defenses. Previous NMS-oriented mitigations cap the candidates entering NMS~\cite{chen2023overload}, resize inputs~\cite{schoof2024beyond}, or adversarially train the detector to suppress phantoms under a hardware-derived box budget~\cite{underload}, whereas NMS-free pipelines require bounding the workload entering downstream association. The defended component therefore shifts together with the attack surface.

\subsection{Defense: bounded-admission layer}
\label{sec:defense}
Our analysis points to an architecture-agnostic mitigation---a \emph{bounded
admission layer} between detector and tracker. A global cap $K$ blocks the flood
mode, and a per-region cap $K_r$ over a spatial grid blocks the spread mode (a plain
global cap could be filled by scattered phantoms). Unlike raising the detection
threshold $\tau$ alone---which SlowTrack reports does not defend its attack and
degrades benign recall~\cite{ma2024slowtrack}---the caps bound the number of
admitted detections by construction, regardless of how the attacker distributes
confidence mass; we did not experimentally compare against threshold tuning.
The cap bounds the \emph{per-frame input} to association: with at most $K$
detections admitted per frame, the active-plus-buffered track pool can still
accumulate to roughly $K\times\text{track\_buffer}$ over sustained attack, so the
layer limits the admission rate rather than proving a constant worst-case
\texttt{update()} bound. TrackShield~\cite{gu2026trackshield} formalizes this
distinction for SlowTrack-style pool inflation: per-stage admission caps inside
the tracker bound the admitted association dimensions and the per-frame pool
inflow for every input, whereas a constant pool bound additionally requires a
retention rule that ages every unrefreshed track, and in practice pruning of
likely phantom tracks is needed to keep the pool near its benign level. Empirically it restores the end-to-end multiplier to
$1.01$--$1.03\times$ (Table~\ref{tab:defense_jetson}). The layer runs before
the tracker and adds negligible latency---a confidence sort plus $O(D)$ grid
bucketing over the $D$ detections, orders of magnitude below the
millisecond-scale tracker \texttt{update()}---and is both detector- and
tracker-agnostic.

\begin{table}[htbp]
\centering
\begin{small}
\caption{Bounded-admission defense under the universal
$\epsilon=8/255$ attack (Jetson AGX Orin, TensorRT FP16;
ByteTrack). ``Adv. mult.'' denotes the end-to-end latency
multiplier. Detection retention is the fraction of clean
detections admitted after applying the admission cap.}
\label{tab:defense_jetson}
\setlength{\tabcolsep}{5pt}
\begin{tabular}{lccc}
\multicolumn{4}{c}{\emph{(a) RT-DETR-l (query-based, high output density)}}\\
\toprule
Defense & \shortstack{Clean det.\\admitted} & \shortstack{Detection\\retention} & \shortstack{e2e\\mult.} \\
\midrule
None (uncapped)  & 22.8 & 1.00 & 1.08$\times$ \\
$K{=}40$         & 22.1 & 0.97 & 1.07$\times$ \\
$K{=}20$         & 16.0 & 0.70 & 1.06$\times$ \\
$K{=}10$         & \phantom{0}9.6 & 0.42 & \textbf{1.01$\times$} \\
$K_r{=}3$ (grid) & \phantom{0}7.4 & 0.33 & 1.04$\times$ \\
\midrule
\multicolumn{4}{c}{\emph{(b) YOLO26-n (one-to-one, lower output density)}}\\
\midrule
None (uncapped)  & 6.0 & 1.00 & 1.17$\times$ \\
$K{=}15$         & 6.0 & 1.00 & 1.16$\times$ \\
$K{=}10$         & 5.7 & 0.95 & 1.10$\times$ \\
$K{=}5$          & 4.0 & 0.67 & \textbf{1.03$\times$} \\
$K_r{=}3$ (grid) & 5.3 & 0.88 & 1.09$\times$ \\
\bottomrule
\end{tabular}
\end{small}
\end{table}

Table~\ref{tab:defense_jetson} reports the on-device security--utility tradeoff
under the universal $8/255$ attack. Its ``Adv.\ mult.''\ is the \emph{end-to-end}
multiplier, consistent with the uncapped end-to-end multipliers in Table~\ref{tab:eps_jetson_full} (the small differences, e.g.\ $1.09\times$ vs.\ $1.08\times$, reflect capture-to-capture timing variance between separate on-device runs with the same frozen delta), \emph{not} the tracker-only $2.22\times$ of Table~\ref{tab:univ}(a). The cap acts on the tracker stage as the thesis predicts. For the
high-density query head (RT-DETR-l) it lowers the end-to-end multiplier from
$1.08\times$ to $1.01\times$ at $K{=}10$ while retained detections fall to
$0.42$: association never sees more than $K$ detections whatever flood a denser
scene achieves. For the low-density one-to-one head (YOLO26-n) the cap cuts the multiplier from $1.17\times$ to
$1.03\times$ at $K{=}5$ ($0.67$ retained), at negligible cost until $K$ drops
below the typical clean detection count. The per-region cap $K_r{=}3$ reaches
the strongest tracker bound (RT-DETR-l down to $7.4$ admitted detections) by
directly limiting the \emph{spread} mode that a global cap alone cannot, at a
larger utility cost. %

\section{Related Work}
\label{sec:related}
Our work intersects five lines of research: availability attacks on deep
learning in general, detector-side and pipeline-level latency attacks on
perception, integrity attacks on multi-object tracking, defenses against latency
attacks, and fixed-computation NMS-free detection. Recent surveys organize the
first two by attack surface and system impact~\cite{gu2026survey,meftah2025energy}.

\textbf{Availability attacks on deep learning.} Sponge
examples~\cite{shumailov2021sponge} first showed that crafted inputs can raise
the energy and latency of deployed networks. Later attacks target
content-dependent computation wherever it appears: activation sparsity exploited
by accelerators~\cite{krithivasan2020sparsity} and spiking
networks~\cite{krithivasan2022snn}, early-exit and other input-adaptive
networks~\cite{haque2020ilfo,hong2021panda}, adaptive token pruning in vision
transformers~\cite{navaneet2024slowformer,yehezkel2024desparsify}, and
autoregressive decoding in translation, captioning, and vision-language
models~\cite{chen2022nmtsloth,chen2022nicgslowdown,gao2024verbose}; sponge
poisoning moves the attack to training time~\cite{cina2025sponge}. We share
their premise that an input can control how much work a deployed stage
performs, but in our pipeline the network is fixed-cost and the
content-dependent stage is the tracker behind it.

\textbf{Latency attacks on perception pipelines.} For camera-based object detection,
Daedalus~\cite{wang2022daedalus}, Phantom Sponges~\cite{shapira2023},
Overload~\cite{chen2023overload}, and later
refinements~\cite{schoof2024beyond} flood NMS with phantom candidates.
SlowTrack~\cite{ma2024slowtrack} extends the payload to tracking in NMS-based
camera pipelines, and Detstorm~\cite{muller2025detstorm} realizes phantom-object
floods physically with projected light. Beyond cameras,
SlowLiDAR~\cite{liu2023slowlidar} slows LiDAR detection pipelines and
CP-FREEZER~\cite{wang2025cpfreezer} attacks cooperative perception through
V2X messages. The EVADE framework~\cite{monteuuis2025evade} cautions that
several published NMS latency attacks lose most of their effect when re-evaluated
across hardware platforms, model formats, and quantization, and that the
resulting increases often stay within downstream latency requirements; this is
one reason we measure every number on the deployment target and separate the
modest $8/255$ regime from the $32/255$ stress test. Execution-time variability
of post-processing is also a timing side channel that can aid other
attacks~\cite{biton2023variable}. These attacks rely on a content-dependent detector-side stage;
we study the case in which the detector no longer has one.

\textbf{Attacks on multi-object tracking.} Integrity attacks on trackers
hijack or suppress tracks by moving detections across
frames~\cite{jia2020fooling,muller2022hijacking}. Our attack leaves identities
aside and targets the tracker's execution time instead.

\textbf{Defenses.} Detector-side mitigations cap the number of candidates
entering NMS~\cite{chen2023overload}, resize inputs~\cite{schoof2024beyond}, or
adversarially train the detector to suppress phantom objects under a
hardware-derived budget on candidate boxes~\cite{underload}. The last work also
reports that DETR-family detectors show no detector-side latency growth with the
number of objects, which is consistent with our finding that the surface moves
downstream rather than disappearing. At the tracker layer, TrackShield~\cite{gu2026trackshield} defends
ByteTrack-style pipelines against SlowTrack with anomaly-triggered birth capping,
phantom-track pruning, and always-on per-stage association caps whose admission
bound holds independently of its monitor. Our bounded-admission layer applies the same admission-control
principle~\cite{buttazzo2011hard} at a complementary point, the detector--tracker
boundary. Because it acts before the tracker, it needs no changes to tracker
internals and is tracker-agnostic by construction, requiring only the detection
list; it also bounds, by construction, per-detection work that some trackers
perform before association, such as appearance-ReID embedding, which an
in-tracker guard does not cover~\cite{gu2026trackshield}; and its per-region cap
additionally targets the spatial spread mode of the flood. The two mechanisms
compose naturally: boundary admission limits what enters the tracker, while
in-tracker guards and pruning limit the persistent state that accumulates
inside it.

\textbf{NMS-free detection.} Our work differs from these works in three aspects. First, we target NMS-free detector families,
including one-to-one detectors (YOLOv10~\cite{wang2024yolov10},
\mbox{YOLO26~\cite{jocher2026yolo26}}) and query-based detectors
(RT-DETR~\cite{zhao2024rtdetr}): attacks whose latency payload relies
specifically on content-dependent NMS do not directly transfer to an NMS-free
inference path, though other detector-side mechanisms may remain.
Second, we recover differentiable confidence tensors from both detector
families, enabling a unified overload attack. Third, we provide a comparative
on-device evaluation showing that susceptibility is governed primarily by the achievable
detection flood rather than detector head type alone.

\section{Limitations and Conclusion}

\textbf{Threats to validity.}
(1)~\emph{Threat model.} Our attack assumes a digital white-box adversary who
can alter pixels between the camera and the detector (e.g., via a compromised
ISP, camera driver, or uplink, or by remote optical injection into the
sensor~\cite{man2020ghostimage}) and who knows the deployed model; physically
realizable variants (patches, stickers, or projected
light~\cite{muller2025detstorm,cao2021invisible}) remain future work.
(2)~\emph{Utility metric.} We quantify defense utility only as detection-count
retention, a coarse proxy: we do not report AP, recall, MOTA~\cite{bernardin2008clear},
IDF1~\cite{ristani2016idf1}, or HOTA~\cite{luiten2021hota} under
attack or defense, so our results do not establish that perception quality is
preserved---only that the count of admitted detections is bounded.
(3)~\emph{Non-adaptive evaluation.} The bounded-admission layer is evaluated
only against our non-adaptive attack; an adversary aware of $K$ could in
principle optimize phantom confidences to out-rank true detections, so the
layer bounds workload, not integrity. Admission inside the tracker, as in
TrackShield~\cite{gu2026trackshield}, would additionally bound per-stage
association work and pool inflow against such an adversary, but we have not
evaluated it on NMS-free pipelines.
(4)~\emph{Execution model.} All timing uses a sequential detector-then-tracker
execution; in a pipelined deployment the effect depends on which stage is the bottleneck: while the tracker stays shorter than the detector, its overload is absorbed by pipeline slack, but once it becomes the longest stage the frame period grows with the tracker itself, so the throughput loss can approach the tracker-only multiplier rather than our sequential end-to-end multiplier. Our end-to-end numbers are therefore neither an upper nor a lower bound for pipelined systems.
(5)~\emph{Dataset and power.} We evaluate on KITTI with four held-out
sequences; sequence-level statistical power is limited (even if all four clips
show a multiplier above 1, a one-sided sign test reaches only $p{=}0.0625$),
and the 593-frame pool is a dense subset, so pooled aggregates weight dense
scenes.
(6)~\emph{Class filtering.} The floods place many phantoms in COCO classes irrelevant to driving (Fig.~\ref{fig:y26} shows chairs, umbrellas, and a dining table); a pipeline that filters classes before tracking would remove part of the present flood, and we have not evaluated an attack restricted to driving-relevant classes.
(7)~\emph{Missing ablations and baselines.} We do not report ablations of the flood, spread, and margin terms, a comparison against an adapted SlowTrack objective, or confidence intervals for most timing entries; the role of the spread term is argued, not isolated experimentally.
(8)~\emph{Class-agnostic retention.} Count-based retention does not distinguish
safety-critical classes; class-aware caps (e.g., safety-class exemptions) would
be needed to protect vulnerable road users but introduce their own gaming
surface.

Removing Non-Maximum Suppression (NMS) eliminates the classical
detector-side latency bottleneck but relocates the availability attack
surface downstream to multi-object tracking. Our Jetson AGX Orin
evaluation shows that the tracker overload induced by TrackFlood is governed
primarily by the achievable detection flood rather than detector head
type, and that a lightweight bounded-admission layer bounds tracker
workload at a quantified, non-trivial utility cost.

\textbf{Implications for real-time system design.}
The worst-case execution time of a perception pipeline stage is, under this
threat, attacker-controllable through its input content: schedulability
analysis that treats tracker cost as a bounded constant, with a worst-case
execution time derived from benign inputs~\cite{wilhelm2008wcet}, is unsafe, and
per-frame deadline guarantees must account for adversarial input
distributions, much as mixed-criticality scheduling already distinguishes
execution-time assumptions of different assurance levels~\cite{vestal2007mixed}.
Classical admission control~\cite{buttazzo2011hard} at the detector--tracker boundary
is an effective first-line mitigation, restoring a per-frame admission bound
at negligible overhead. Real-time perception stacks should therefore
evaluate timing at the system level rather than the detector level alone,
and co-design detector, tracker, and admission control when developing
latency-aware defenses and schedulers.

\bibliographystyle{ACM-Reference-Format}
\bibliography{references}

\end{document}